\documentclass[amsmath,floatfix,twocolumn,superscriptaddress]{revtex4-1}
\usepackage{subfigure}
\usepackage{siunitx}
\usepackage{amssymb}
\usepackage{amsmath}
\usepackage{graphicx}
\usepackage{array}
\usepackage{dcolumn}
\usepackage{psfrag}
\usepackage{bm}
\usepackage{color}
\usepackage{multirow}
\usepackage{gensymb}
\usepackage{mathptmx}

\begin{document}

\title{Molecular origins of colossal barocaloric effects in plastic crystals}

\author{Ares Sanuy}
 \affiliation{Group of Characterization of Materials, Departament de F\'{i}sica, Universitat Polit\`{e}cnica de Catalunya,
    Campus Diagonal-Bes\`{o}s, Av. Eduard Maristany 10--14, 08019 Barcelona, Spain}
 \affiliation{Barcelona Research Center in Multiscale Science and Engineering, Universitat Politècnica de Catalunya,
    Campus Diagonal-Bes\`{o}s, Av. Eduard Maristany 10--14, 08019 Barcelona, Spain}

\author{Carlos Escorihuela-Sayalero}
 \affiliation{Group of Characterization of Materials, Departament de F\'{i}sica, Universitat Polit\`{e}cnica de Catalunya,
    Campus Diagonal-Bes\`{o}s, Av. Eduard Maristany 10--14, 08019 Barcelona, Spain}
 \affiliation{Barcelona Research Center in Multiscale Science and Engineering, Universitat Politècnica de Catalunya,
    Campus Diagonal-Bes\`{o}s, Av. Eduard Maristany 10--14, 08019 Barcelona, Spain}

\author{Pol Lloveras}
 \affiliation{Group of Characterization of Materials, Departament de F\'{i}sica, Universitat Polit\`{e}cnica de Catalunya,
    Campus Diagonal-Bes\`{o}s, Av. Eduard Maristany 10--14, 08019 Barcelona, Spain}
 \affiliation{Barcelona Research Center in Multiscale Science and Engineering, Universitat Politècnica de Catalunya,
    Campus Diagonal-Bes\`{o}s, Av. Eduard Maristany 10--14, 08019 Barcelona, Spain}

\author{Josep-Lluís Tamarit}
 \affiliation{Group of Characterization of Materials, Departament de F\'{i}sica, Universitat Polit\`{e}cnica de Catalunya,
    Campus Diagonal-Bes\`{o}s, Av. Eduard Maristany 10--14, 08019 Barcelona, Spain}
 \affiliation{Barcelona Research Center in Multiscale Science and Engineering, Universitat Politècnica de Catalunya,
    Campus Diagonal-Bes\`{o}s, Av. Eduard Maristany 10--14, 08019 Barcelona, Spain}

\author{Luis Carlos Pardo}
 \email{luis.carlos.pardo@upc.edu}
 \affiliation{Group of Characterization of Materials, Departament de F\'{i}sica, Universitat Polit\`{e}cnica de Catalunya,
    Campus Diagonal-Bes\`{o}s, Av. Eduard Maristany 10--14, 08019 Barcelona, Spain}
 \affiliation{Barcelona Research Center in Multiscale Science and Engineering, Universitat Politècnica de Catalunya,
    Campus Diagonal-Bes\`{o}s, Av. Eduard Maristany 10--14, 08019 Barcelona, Spain}

\author{Claudio Cazorla}
 \email{claudio.cazorla@upc.edu}
 \affiliation{Group of Characterization of Materials, Departament de F\'{i}sica, Universitat Polit\`{e}cnica de Catalunya,
    Campus Diagonal-Bes\`{o}s, Av. Eduard Maristany 10--14, 08019 Barcelona, Spain}
 \affiliation{Barcelona Research Center in Multiscale Science and Engineering, Universitat Politècnica de Catalunya,
    Campus Diagonal-Bes\`{o}s, Av. Eduard Maristany 10--14, 08019 Barcelona, Spain}

\begin{abstract}
In recent years, orientationally disordered crystals, or plastic crystals, have transformed the field of 
solid-state cooling due to the significant latent heat and entropy changes associated with their temperature-induced 
molecular order-disorder phase transition, which can produce colossal caloric effects under external field stimuli. However, 
the molecular mechanisms underlying these huge caloric effects remain inadequately understood, and general principles 
for enhancing the performance of caloric plastic crystals are lacking. Previous studies have predominantly focused on 
molecular rotations, overlooking other potentially critical factors, such as lattice vibrations and molecular conformations. 
In this study, we employ classical molecular dynamics (MD) simulations to both replicate and elucidate the microscopic 
origins of the experimentally observed colossal barocaloric (BC) effects --those driven by hydrostatic pressure-- in the 
archetypal plastic crystal neopentyl glycol (NPG). Our MD simulations demonstrate that in NPG, the combined BC response 
and phase-transition entropy changes arising from lattice vibrations and molecular conformations are nearly equal to those 
from molecular reorientations, contributing $45$\% and $55$\%, respectively. These findings suggest that, alongside hydrogen 
bonding --which directly impacts molecular rotational dynamics-- lattice vibrational and molecular structural features, 
often overlooked, must be integrated into the rational design and modeling of advanced caloric plastic crystals. These 
insights are not only of significant fundamental interest but also essential for driving the development of next-generation 
solid-state refrigeration technologies.
\\

\end{abstract}

\maketitle

\section{Introduction}
\label{sec:intro}
Solid-state cooling methods offer energy-efficient and environmentally friendly alternatives to conventional refrigeration 
technologies, which rely on compression cycles of greenhouse gases \cite{manosa13,manosa2010,manosa17,cazorla19,lloveras21,
Hou2022,rurali24}. Caloric materials, when subjected to moderate external field variations (e.g., magnetic, electric, 
mechanical or light irradiation), exhibit significant adiabatic temperature changes ($|\Delta T| \sim 1$--$10$~K) due to 
phase transformations that involve substantial isothermal entropy changes ($|\Delta S| \sim 10$--$100$~J~K$^{-1}$~kg$^{-1}$). 
Solid-state cooling leverages these caloric effects to design efficient refrigeration cycles.

From a practical standpoint, achieving responsive and efficient thermal devices under repeated application and removal 
of driving fields requires both large and reversible $|\Delta T|$ and $|\Delta S|$. Among the various caloric effects, 
mechanocaloric responses induced by uniaxial stress (elastocaloric) and hydrostatic pressure (barocaloric, BC) are 
particularly promising due to their substantial figures of merit \cite{aznar17,cazorla16,sagotra17,sagotra18,manosa20,li24}.

Recently, colossal and reversible BC effects ($|\Delta S| \ge 100$~J~K$^{-1}$~kg$^{-1}$) have been observed in several 
families of plastic crystals undergoing molecular order-disorder (OD) phase transitions under pressure changes of the order 
of $0.1$~GPa. Prominent examples include neopentane derivatives \cite{lloveras19,li19,cazorla19b,aznar20}, adamantane-based 
compounds \cite{aznar21,salvatori22}, carboranes \cite{li22}, and closo-borate crystals \cite{sau21,zeng24}. These materials 
constitute a pivotal class of disordered systems driving leading-edge advancements in barocaloric refrigeration technology
\cite{li24,cirillo22}. However, a complete and fundamental understanding of the physical mechanisms behind the colossal 
BC effects observed in plastic crystals remains elusive. 

\begin{figure*}[t]
\includegraphics[width=1.0\linewidth]{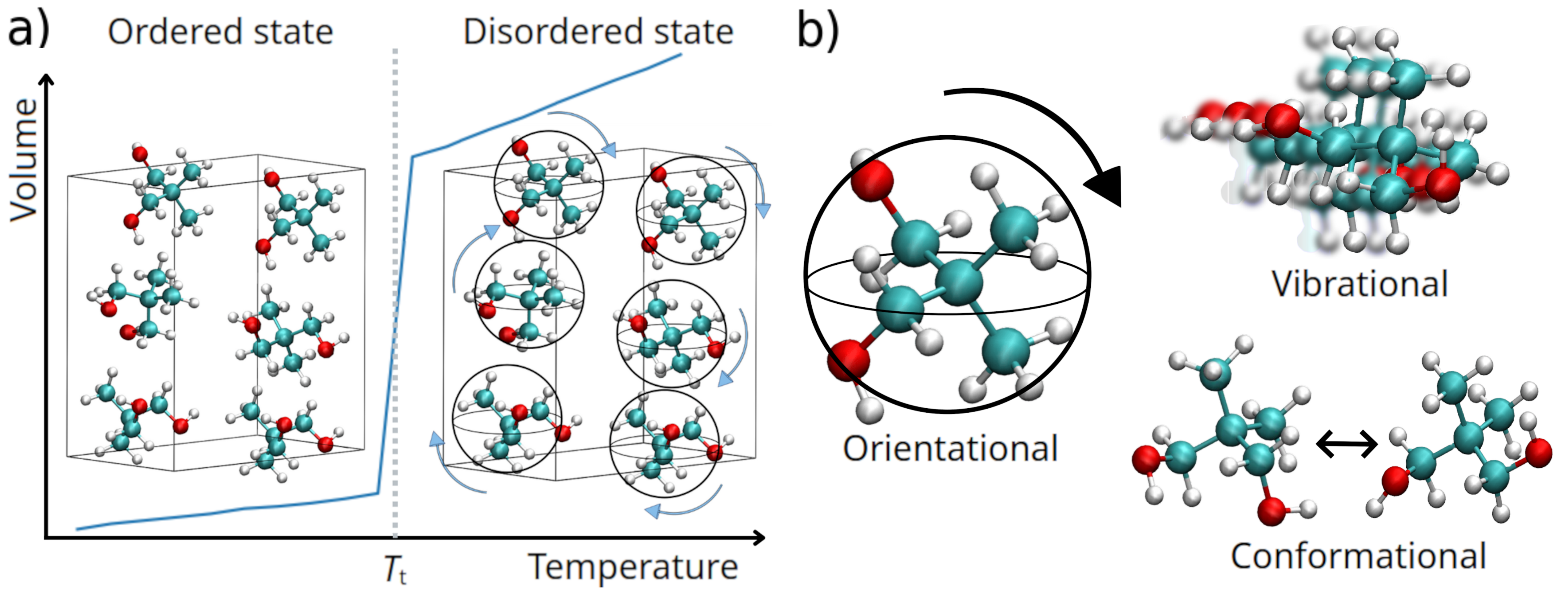}
        \caption{{\bf Sketch of the $T$-induced OD phase transition in NPG and different sources of entropy in 
	the high--$T$ disordered state.} (a)~Below the transition temperature, $T_{t}$, NPG presents an ordered phase in 
	which only molecular vibrations and lattice phonons contribute to its entropy. Above $T_{t}$, NPG becomes orientationally 
	disordered and, besides molecular vibrations and lattice phonons, the angular degrees of freedom, including possible 
	molecular orientations and conformations, also contribute to the system entropy. The OD phase transition 
	is accompanied by a large change in volume and entropy. (b)~Representation of the three possible sources of entropy 
	in the high-$T$ disordered phase of NPG.}
\label{fig1}
\end{figure*}

To date, the large $|\Delta T|$ and $|\Delta S|$ measured in plastic crystals have been primarily attributed to substantial 
entropy changes arising from molecular orientational disorder \cite{hui22,li20,oliveira23,marin24}. This molecular disorder 
emerges at high temperatures and typically is accompanied by an enormous volume increase of the order of $10$\%
\cite{lloveras19,li19,cazorla19b,aznar20} (Fig.~\ref{fig1}a). Notably, hydrogen bond interactions between molecules have 
been found to play a crucial role in the orientational dynamics, hindering or facilitating the molecular reorientations 
depending on whether these interactions are reinforced or weakened \cite{hui22,li20,tamarit97}. As a result, the origin of 
colossal BC effects has been predominantly linked to orientational dynamics and hydrogen bonding, often overlooking other 
potential sources of entropy such as lattice vibrations and molecular conformations (Fig.~\ref{fig1}b).

Interestingly, recent experimental and theoretical studies have highlighted the significant contributions of vibrational 
and molecular conformational degrees of freedom to the phase-transition entropy change in plastic crystals and related 
orientationally disordered materials, such as hybrid organic-inorganic perovskites \cite{zeng24,escorihuela24,cazorla24,
meijer23,yuan22,walker24}. For instance, approximately $60$\% of the reversible $|\Delta S|$ measured in adamantane has 
been attributed solely to vibrational effects \cite{meijer23}. Similarly, in the case of the mono-carba-closo-dodecaborate 
LiCB$_{11}$H$_{12}$, only about $30$\% of the phase transition entropy change is associated with molecular reorientations 
\cite{zeng24}. Moreover, the conformational flexibility and soft nature of hydrocarbon chains, which enable numerous 
degenerate molecular configurations in the high-$T$ disordered phase, have been identified as the primary mechanisms driving 
substantial BC effects in two-dimensional metal-halide perovskites \cite{seo22,li21} and hybrid organic-inorganic halide 
salts \cite{seo24}.

These findings suggest that molecular reorientations alone might not fully account for the colossal BC effects measured 
in plastic crystals. This possibility underscores the urgent need to rigorously identify and quantify all potential 
sources of entropy change in such materials. Addressing this challenge is crucial for two primary reasons. First, 
to validate and potentially improve the phenomenological free-energy models currently used to describe OD phase transitions 
in molecular crystals \cite{oliveira23,marin24}. And second, to accelerate the rational design and development of new 
energy-efficient solid-state cooling materials, thus paving the way for their broader technological adoption.

In this study, we employed classical molecular dynamics (MD) simulations to uncover the microscopic origins of the experimentally 
observed colossal BC effects in the archetypal plastic crystal neopentyl glycol (CH$_{3}$)$_{2}$C(CH$_{2}$OH)$_{2}$ (NPG). 
Our simulations reveal that the combined contributions of lattice vibrations and molecular conformational changes to the 
phase-transition entropy change and BC descriptor $\Delta S$ are comparable in magnitude to those from molecular reorientations.  
These findings underscore the importance of incorporating not only molecular rotational dynamics but also structural features 
and lattice phonons into the microscopic understanding and smart engineering of caloric plastic crystals. 

The structure of this article is as follows. First, we provide an overview of the employed simulation and entropy evaluation 
methods, which have been recently introduced in works \cite{zeng24,escorihuela24,cazorla24}. Next, we present our results  
for the OD phase transition and BC effects in NPG, discussing their alignment with the reported experimental data. This is 
followed by a detailed analysis of the different contributions to the calculated entropy change and BC performance. Finally, 
we summarize our main findings in the conclusions section.

\section{Computational Approach}
\label{sec:computation}

\subsection{Simulations overview and general definitions}
\label{subsec:general}
We conducted comprehensive classical molecular dynamics (MD) simulations in the $NPT$ ensemble for NPG across a wide
range of temperature ($T$) and pressure ($P$) conditions. Our $NPT$--MD simulations employed a CHARMM-type force 
field comprising bonded (e.g., angle bending, bond stretching and dihedral torsion) and nonbonded (i.e., van der Waals 
and electrostatic) interactions \cite{charmm,swissparam1,swissparam2}. This classical interatomic potential accurately 
captures the physical behavior of NPG at low and high temperatures (Sec.~\ref{sec:results}). Further details of our 
$NPT$--MD simulations can be found in the Methods section.

The total entropy of the low-$T$ ordered and high-$T$ disordered phases of NPG, $S$, were determined as a function 
of pressure and temperature using the relation:
\begin{equation}
	S(P,T) = S_{\rm vib}(P,T) + S_{\rm ang}(P,T)~, 
\label{eq:stot}
\end{equation}
where $S_{\rm vib}$ is the entropy contribution from the lattice vibrations and $S_{\rm ang}$ from the molecular 
angular degrees of freedom. 

Likewise, the molecular angular entropy was evaluated with the expression:
\begin{equation}
	S_{\rm ang}(P,T) = S_{\rm ori}(P,T) + S_{\rm conf}(P,T)~,
\label{eq:sang}
\end{equation}
where $S_{\rm ori}$ is the entropy contribution from the orientational degrees of freedom, and $S_{\rm conf}$ 
from the molecular conformational degrees of freedom. Molecule-molecule orientational correlations have been 
neglected throughout this work since their contribution to the total entropy is expected to be of the order of 
$10$~J~K$^{-1}$~kg$^{-1}$ \cite{escorihuela24}, that is, an order of magnitude smaller than $S_{\rm ori}$ and 
$S_{\rm conf}$ (Sec.~\ref{sec:discussion}). 

In this study, we primarily have focused on reproducing and understanding at the molecular level the origins of the 
experimental colossal BC effects measured in NPG. In this context, a physical quantity of interest is the OD 
phase-transition entropy change, defined as:
\begin{equation}
	\Delta S_{t} = \Delta S_{\rm vib} + \Delta S_{\rm ori} + \Delta S_{\rm conf}~,
\label{eq:deltast}
\end{equation}
where $\Delta S_{\rm x} \equiv S_{\rm x}^{\rm disord} - S_{\rm x}^{\rm ord}$ and all the entropy terms are evaluated at 
the phase-transition temperature, $T_{t}$, corresponding to a fixed pressure. The main improvement of this $\Delta S_{t}$ 
definition, compared to those in previous works \cite{zeng24,cazorla24}, is the inclusion of molecular conformational 
changes ($\Delta S_{\rm conf}$).

\begin{figure*}[t]
\includegraphics[width=1.0\linewidth]{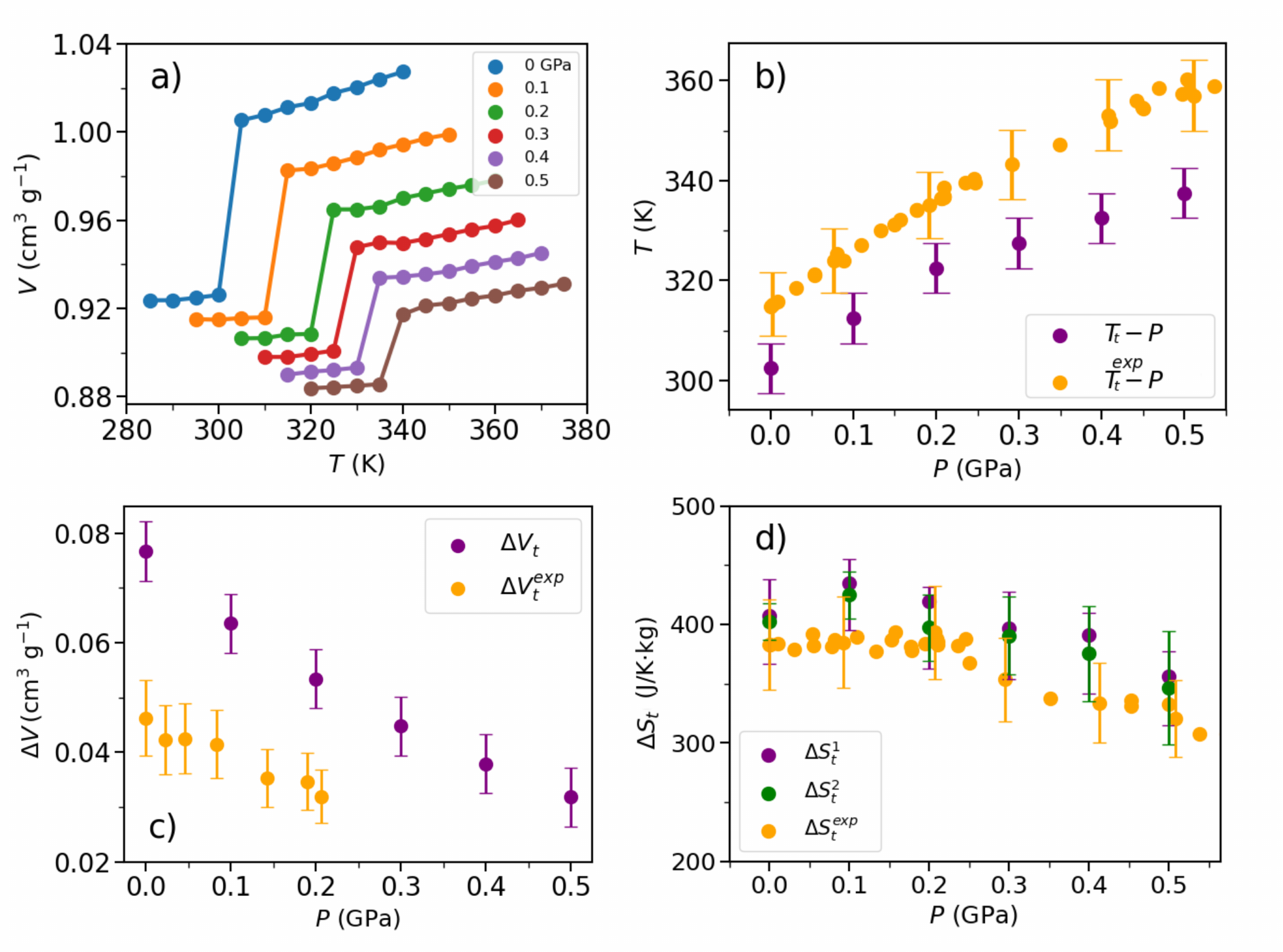}
       \caption{{\bf Simulated OD phase transition in NPG expressed as a function of pressure and temperature.}
        (a)~Volume evolution obtained from the $NPT$--MD simulations. Solid lines are guides for the eye.
        (b)~Transition temperature expressed as a function of pressure (violet dots); experimental data (orange dots)
        are shown for comparison.
        (c)~Transition volume change expressed as a function of pressure (violet dots); experimental data (orange dots)
        are shown for comparison.
        (d)~Transition entropy change expressed as a function of pressure (violet and green dots). $\Delta S_{t}^{1}$
        and $\Delta S_{t}^{2}$ were estimated indirectly using the Clausius-Clapeyron relation and directly from the
        $NPT$--MD simulations, respectively. Experimental data (orange dots) are shown for comparison.
       }
\label{fig2}
\end{figure*}

\subsection{Vibrational entropy}
\label{subsec:vibrational}
The vibrational density of states $\rho (\omega)$, where $\omega$ represents lattice vibration frequency, provides information 
on the phonon spectrum of a crystal and allows to estimate key thermodynamic quantities like the vibrational free energy, 
$F_{\rm vib}$, and vibrational entropy, $S_{\rm vib} = -\frac{\partial F_{\rm vib}}{\partial T}$. A possible manner 
of calculating $\rho (\omega)$ from the outputs of $NPT$--MD simulations consists in estimating the Fourier transform of 
the velocity autocorrelation function (VACF) \cite{cazorla24,sagotra19,cibran23}, defined like:
\begin{equation}
    \text{VACF}(t) = \left\langle \mathbf{v}(0)\cdot \mathbf{v}(t) \right\rangle = \frac{1}{N} \sum_i^{N} \mathbf{v_i}(0)\cdot\mathbf{v_{i}}(t)~,
    \label{eq:vacf}
\end{equation}
where $\mathbf{v_i}(t)$ represents the velocity of the $i$-th particle and $\langle \cdots \rangle$ statistical average 
performed in the $NPT$ ensemble. Subsequently, the vibrational density of states can be estimated like: 
\begin{equation}
    \rho(\omega) = \int_0^{\infty}\text{VACF}(t)~e^{i\omega t}~dt~,
    \label{eq:vdos}
\end{equation}
which fulfills the normalization condition:
\begin{equation}
\int_0^{\infty} \rho(\omega)~d\omega = 3N~,
\label{eq:vdos-norm}
\end{equation}
$3N$ being the number of real and positively defined phonon frequency branches of the crystal. 

Upon determination of $\rho$, the vibrational free energy can be straightforwardly estimated as a function of pressure
and temperature with the formula 
\cite{cazorla17}:
\begin{eqnarray}
	F_{\rm vib} (P,T) & = & k_{B} T \times \nonumber \\  
	& & \int_0^{\infty} \ln{\left[ 2\sinh{\left(\frac{\hbar\omega}{2k_BT}\right)}\right]} \rho(\omega)~d\omega~,
\label{eq:Fvib}
\end{eqnarray}
where $k_{B}$ is the Boltzmann's constant and $\rho(\omega)$ depends both on $P$ and $T$. 
Consistently, the vibrational entropy adopts the expression:
\begin{eqnarray}
S_{\rm vib} (P,T) & = & -\int_0^{\infty} k_{B}\ln{\left[2\sinh{\left(\frac{\hbar\omega}{2k_BT}\right)}\right]} \rho(\omega)~d\omega + \nonumber \\ 
& & \int_0^{\infty} \frac{\hbar\omega}{2T} \tanh^{-1}{\left(\frac{\hbar\omega}{2k_BT}\right)} \rho(\omega)~d\omega~.
\label{eq:svib}
\end{eqnarray}

\subsection{Molecular orientational entropy}
\label{subsec:orientational}
For a continuous random variable $x$ with probability density $f(x)$, its entropy is defined as \cite{informationtheory}:
\begin{equation}
S = - \int_{X} f(x) \log{f(x)}~dx~,
\label{eq:shanon}
\end{equation}
where the integral runs over all possible values of $x$. If $x$ represents a microstate characterizing a particular thermodynamic 
macrostate, then the following Gibbs entropy can be defined for the system of interest \cite{pathria1972}:  
\begin{equation}
S_{G} = -k_{B} \int_{X} f(x) \log{f(x)}~dx~.
\label{gibbs-entropy}
\end{equation}
 
In an orientationally disordered crystal, molecules rotate in a quasi-random manner and their orientation with respect
to a static reference system can be described by the three angles $\theta$, $\phi$, and $\psi$ (Sec.~\ref{subsec:ori} 
and Fig.~\ref{fig4}a) \cite{escorihuela24}. By assuming the NPG molecules in the crystal to rotate independently one from 
the other, one can estimate a probability density function for their orientation, $f(\theta, \phi,\psi)$, from the atomistic 
trajectories generated during long $NPT$--MD simulations. In this case, the following three-dimensional molecular orientational 
entropy can be defined, under the implicit assumption that the dimensions of the molecules remain fixed 
\cite{karplus96,pardo16,pardo15}:
\begin{eqnarray}
	S_{\rm ori}(P,T) & = &  S_{\rm ori}^{0}(P,T) -k_{B} \int f(\theta, \phi, \psi) \log{[f(\theta, \phi, \psi)]}  
	\times \nonumber \\
		    & & d\cos(\theta)~d\phi~d\psi~,
\label{eq:sang-cont}
\end{eqnarray}
where $S_{\rm ori}^{0}$ is a reference orientational entropy term. (For a fluid, the value of this reference entropy term 
matches that of an ideal gas under the same temperature and density conditions as the system of interest 
\cite{karplus96,pardo16,pardo15}; however, for an orientationally disordered solid, this reference term is not as 
straightforward to define \cite{cazorla24}.) 

In practice, the calculation of $S_{\rm ori}$ entails the construction of histograms for which the continuous polar variables 
are discretised, $\lbrace \theta, \phi, \psi \rbrace \to \lbrace \theta_{i},\phi_{i}, \psi_{i} \rbrace$. Accordingly, one may 
define the bin probabilities \cite{informationtheory}:
\begin{eqnarray}
	p_{i}(\theta_{i},\phi_{i},\psi_{i}) \approx f(\theta_{i},\phi_{i},\psi_{i}) \times \nonumber  \\
	\Delta\cos(\theta) ~\Delta\phi ~\Delta\psi~,
\label{eq:binprob}       
\end{eqnarray}
where $\Delta\cos(\theta)~\Delta\phi~\Delta\psi$ is the volume of a histogram bin (selected to be constant in this study). 
Consistently, one can rewrite the molecular orientational entropy in the discretised form \cite{zeng24,escorihuela24,
cazorla24,informationtheory}:
\begin{eqnarray}
	\Delta S_{\rm ori}(P,T) & = & -k_{B} \sum_{i} p_{i} \log{p_{i}} + \nonumber \\ 
	               &   &  k_{B} \log{[\Delta\cos(\theta)~ \Delta\phi~ \Delta\psi]}~,
\label{eq:sang-disc}        
\end{eqnarray}
where the value of the reference entropy term in Eq.~(\ref{eq:sang-cont}) has been offset.

\subsection{Molecular conformational entropy}
\label{subsec:conforma}
At finite temperatures, NPG molecules may undergo conformational changes that contribute to the total entropy, 
$S_{\rm conf}$. The dihedral angles $\alpha$ and $\beta$ (Sec.~\ref{subsec:conf} and Fig.~\ref{fig5}a) can be 
used to monitor such conformational molecular changes. Analogously to the case of the orientational entropy, 
a probability density function can be estimated for these molecular dihedral angles from the atomistic trajectories 
generated during long $NPT$--MD simulations, $f(\alpha, \beta)$. Therefore, a bin probability can be defined, 
$p_{i}(\alpha_{i}, \beta_{i}) \approx f(\alpha_{i}, \beta_{i})~\Delta\alpha~ \Delta\beta$, leading to the 
conformational entropy expression:
\begin{equation}
	\Delta S_{\rm conf}(P,T) = -k_{B} \sum_{i} p_{i} \log{p_{i}} + k_{B} \log{(\Delta\alpha~ \Delta\beta)}~,
\label{eq:conforma}
\end{equation}
where $\Delta\alpha~ \Delta\beta$ is the fixed area of a histogram bin, and the corresponding reference entropy 
term is offset, similar to $\Delta S_{\rm ori}$. Without loss of generality and consistent with the observations 
\cite{zanetti61}, the conformational and orientational entropies of the low-temperature ordered phase of NPG are 
assumed to be negligible in this study.

\begin{figure*}[t]
\includegraphics[width=0.8\linewidth]{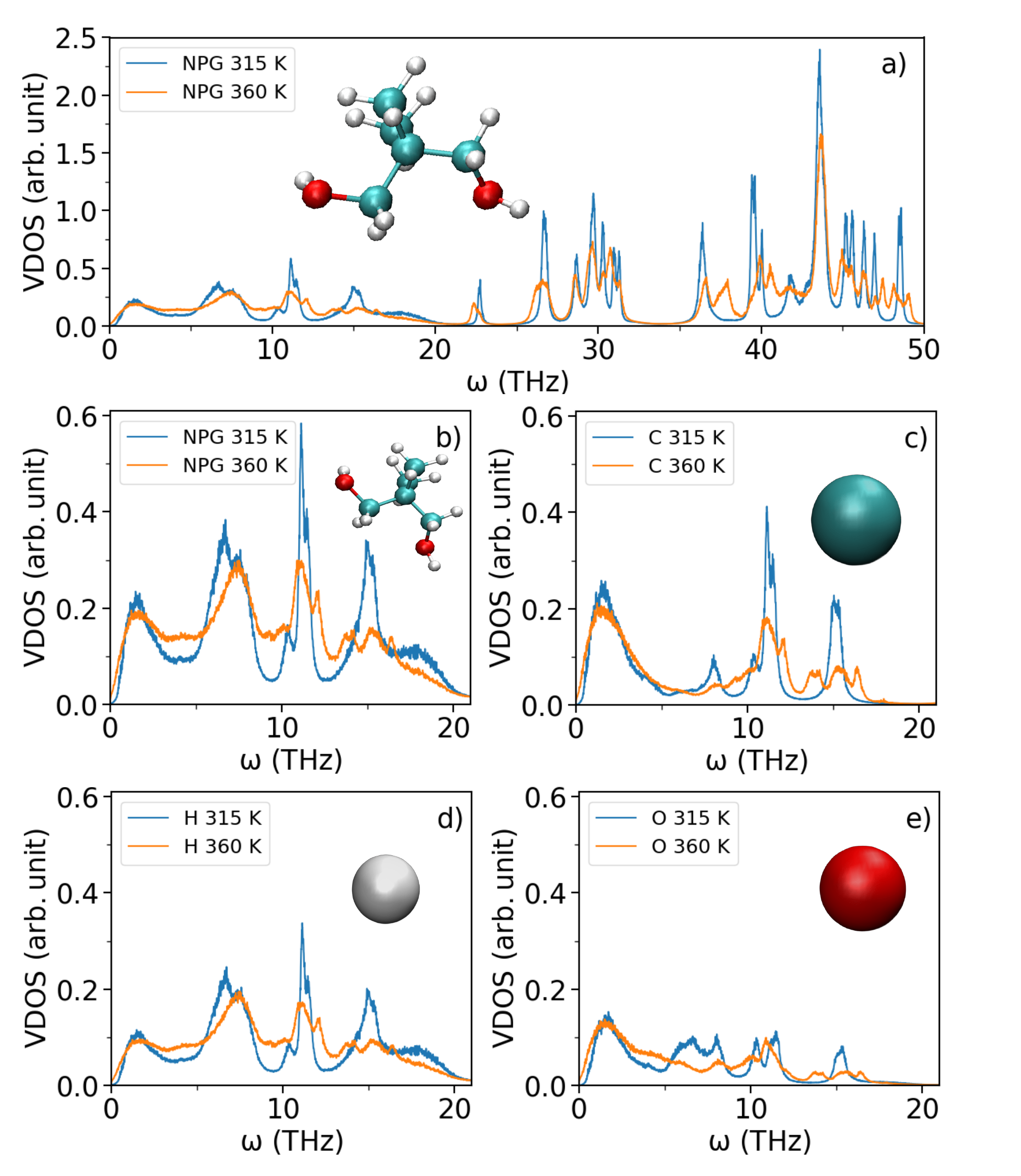}
	\caption{{\bf Vibrational density of states (VDOS) estimated for the ordered and disordered phases of NPG.}
	(a)~Total VDOS calculated for a broad frequency range at zero pressure and temperatures above and below 
	$T_{t}$. (b)~Total VDOS calculated in the low-frequency range. Partial VDOS corresponding to (c)~carbon, 
	(d)~hydrogen and (e)~oxygen atoms. 
       }
\label{fig3}
\end{figure*}

\section{Results}
\label{sec:results}

\subsection{OD phase transition in NPG}
\label{subsec:phasetran}
Figure~\ref{fig2} presents the $P$--$T$ phase diagram of NPG along with key thermodynamic properties derived from 
$NPT$--MD simulations. The low-temperature ordered phase exhibits monoclinic symmetry (space group $P2_{1}/n$) \cite{zanetti61}, 
while the high-temperature disordered phase adopts a cubic (face-centered) lattice, in consistent agreement with 
experimental observations \cite{lloveras19,li19}. The onset of molecular reorientations in the high-temperature 
disordered phase is marked by a significant and abrupt volume increase (Fig.~\ref{fig2}a). For instance, at zero 
pressure, the volume change at the phase transition corresponds to approximately $8.5$\% of the volume of the 
low-temperature ordered phase (Figs.~\ref{fig2}a,c). This relative volume expansion decreases progressively under 
compression (e.g., $\Delta V_{t} / V \approx 3$\% at $P = 0.5$~GPa). It is found that our $NPT$--MD simulations tend 
to overestimate the experimental volume change observed during the OD phase transition (Fig.~\ref{fig2}c). 
While this overestimation is non-negligible within the range of experimental and theoretical uncertainties, it 
becomes less pronounced under increased pressure.

The $P$--$T$ phase diagram of NPG has been estimated up to a pressure of $0.5$~GPa (Fig.~\ref{fig2}b). The slope of 
the calculated phase boundary is positive and nearly constant within the compression range $0 \le P \le 0.2$~GPa, with 
a value of $dT_{t} / dP = 100 \pm 5$~K~GPa$^{-1}$. At higher pressures ($0.3 \le P \le 0.5$~GPa), the induced 
variation of the transition temperature decreases noticeably, reaching approximately $70$~K~GPa$^{-1}$. At pressures 
of $P \le 0.2$~GPa, our $NPT$--MD simulations slightly underestimate the experimental transition temperature upon heating 
by about $10$~K \cite{lloveras19}. However, the slope of the experimental phase boundary upon heating ($dT_{t}^{\rm exp} 
/ dP = 113 \pm 5$~K~GPa$^{-1}$ \cite{lloveras19}) is in excellent agreement with the value predicted by our simulations. 
At higher pressures, the agreement between the experimental and theoretical phase boundaries diminishes, highlighting 
potential limitations of the employed force field under strong compression.

Figure~\ref{fig2}d compares the experimental and calculated transition entropy changes, $\Delta S_{t}$, as a function 
of pressure. The first theoretical estimate, $\Delta S_{t}^{1}$, was derived using the well-known Clausius-Clapeyron 
equation:
\begin{equation}
	\Delta S_{t} = \left(\frac{dT_{t}}{dP}\right)^{-1} \Delta V_{t}~,
\label{eq:cc-formula}
\end{equation}
where $\Delta V_{t}$ represents the phase-transition volume change. This calculation relies on the simulation data shown 
in Figs.~\ref{fig2}a--c. The second theoretical estimate, $\Delta S_{t}^{2}$, was directly obtained from the $NPT$--MD 
simulations by leveraging the Gibbs free-energy equality of the ordered and disordered phases at the phase transition 
temperature. Using this approach, this quantity was computed as $\Delta S_{t}^{2} = \left( \Delta U_{t} + P \Delta V_{t} \right) 
/ T_{t}$, where $U$ denotes the internal energy of the system. Meanwhile, the experimental value, $\Delta S_{t}^{\rm exp}$, 
was determined from calorimetry measurements \cite{lloveras19}.

Several conclusions can be drawn from the $\Delta S_{t}$ results presented in Fig.~\ref{fig2}d. First, the two theoretical 
estimates, $\Delta S_{t}^{1}$ and $\Delta S_{t}^{2}$, are fully consistent with each other, confirming the technical validity 
of our $NPT$--MD simulations. Second, the experimental and theoretical phase transition entropy changes exhibit excellent 
agreement across all the analyzed pressures within their numerical uncertainties. For instance, under ambient pressure, the 
experimental value is $380 \pm 40$~J~K$^{-1}$~kg$^{-1}$ \cite{lloveras19}, while the theoretical estimate is approximately 
$400$~J~K$^{-1}$~kg$^{-1}$. This remarkable consistency between experimental and theoretical values of $\Delta S_{t}$, 
regardless of pressure, provides robust quantitative validation for the subsequent barocaloric analysis presented in 
Sec.~\ref{subsec:baro}.

\begin{figure*}[t]
\includegraphics[width=1.0\linewidth]{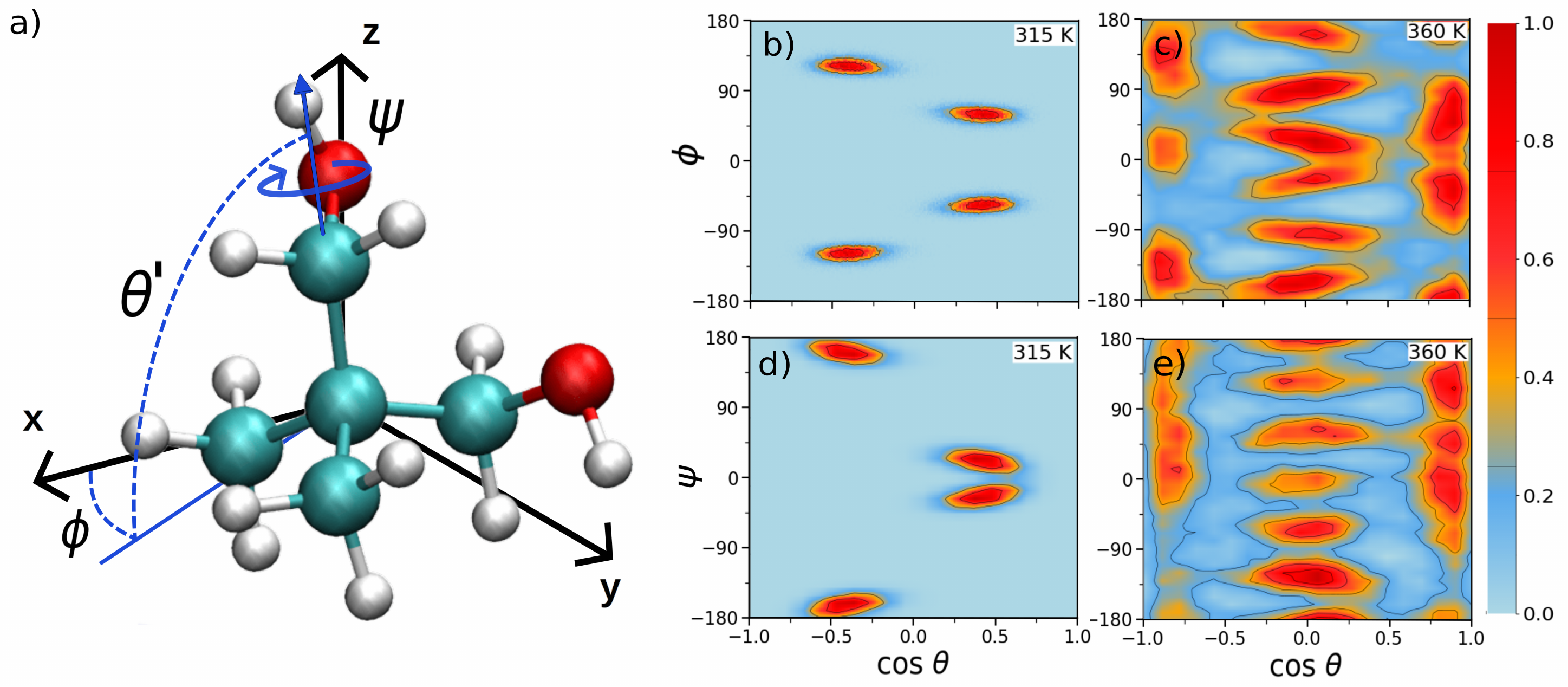}
       \caption{{\bf Molecular orientational degrees of freedom in the ordered and disordered phases of NPG.}
	(a)~Representation of the three angles $\lbrace \theta, \phi, \psi \rbrace$ describing the orientation 
	of NPG molecules in a fixed reference system ($\theta' \equiv \frac{\pi}{2} - \theta $). Black (blue) 
	lines represent the fixed reference system (the co-mobile molecular polar axis and its projection on 
	the fixed $x$--$y$ plane). Average orientational angle histograms obtained from $NPT$--MD simulations 
	at temperatures below (b)--(d) and above (c)--(e) the phase-transition point at zero pressure. 
       }
\label{fig4}
\end{figure*}

\subsection{Vibrational changes across the OD phase transition}
\label{subsec:vib}
Figure~\ref{fig3}a presents the vibrational density of states (VDOS) for the ordered ($T = 315$~K) and disordered 
($T = 360$~K) phases of NPG at zero pressure, as obtained from our $NPT$--MD simulations. The VDOS displays three 
distinct frequency regimes: low ($0 \le \omega \le 20$~THz), medium ($20 \le \omega \le 35$~THz), and high 
($35 \le \omega \le 50$~THz). Across all frequency ranges, the VDOS peaks for the low-$T$ ordered phase are sharper 
and more distinct, reflecting a lower degree of anharmonicity compared to the high-$T$ disordered phase. This heightened 
anharmonicity in the disordered phase contributes to an increased vibrational entropy ($S_{\rm vib}$). In 
Sec.~\ref{sec:discussion}, we present a detailed quantitative analysis of the variation in $S_{\rm vib}$ associated 
with the OD phase transition. Here, we focus on the fundamental vibrational differences between the 
low-$T$ ordered and high-$T$ disordered phases of NPG.  

Near ambient conditions, the most significant frequency interval for evaluating $S_{\rm vib}$ variations is the 
low-$\omega$ range, as it can be inferred from Eq.~(\ref{eq:svib}). Therefore, we focus particularly on the total 
and partial VDOS calculated within the frequency range $0 \le \omega \le 20$~THz (Figs.~\ref{fig3}b--e). 

In the $0 \le \omega \le 5$~THz interval, the high-$T$ disordered phase exhibits a greater accumulation of vibrational 
states than the low-$T$ ordered phase. Specifically, the VDOS surge near $\omega = 0$~THz is more pronounced in the 
disordered phase, and its decline following the first local maximum is less steep (Fig.~\ref{fig3}b). Within this 
frequency range, the vibrational contribution from the C atoms is significantly larger than that of other atoms 
(Figs.~\ref{fig3}c--e). However, the differences in VDOS between the ordered and disordered phases are more prominently 
influenced by the H and O atoms, as the C atom density of states remains largely similar for both phases.

In the $5 \le \omega \le 10$~THz interval, the VDOS of the low-$T$ ordered phase exhibits two distinct peaks, whereas 
only one peak is observed for the high-$T$ disordered phase (Fig.~\ref{fig3}b). Furthermore, the local minima in the VDOS 
are shallower in the disordered phase compared to the ordered phase. Comparisons between the total and partial VDOS 
(Figs.~\ref{fig3}b--e) reveal that the attenuation of vibrational peaks in the disordered phase is primarily driven by 
contributions from the H atoms.

Finally, in the $10 \le \omega \le 20$~THz interval, the high-$T$ disordered phase displays a greater number of VDOS 
peaks compared to the low-$T$ ordered phase, although these peaks are lower in intensity. Comparisons between the total 
and partial VDOS (Figs.~\ref{fig3}b--e) indicate that the emergence of additional vibrational peaks in the disordered 
phase is predominantly influenced by contributions from the C and H atoms.

Notably, our zero-pressure VDOS calculations show fairly good agreement with the available experimental data. Raman 
spectroscopic measurements for the ordered phase of NPG at $T = 296$~K report six distinct peaks in the frequency range 
$5 \leq \omega \leq 20$~THz, located approximately at $7$, $10$, $11$, $12$, $16$, and $17$~THz \cite{granzow95}. Similarly, 
our VDOS calculations at $T = 315$~K also reveal six peaks within the same frequency range, approximately at $7$, $11$, 
$12$, $13$, $15$, and $18$~THz (Fig.~\ref{fig3}b). These findings confirm that our $NPT$--MD simulations provide a 
realistic description of the vibrational properties of NPG, at least for the low-$T$ ordered phase, within the frequency 
range that is relevant to this study.

\begin{figure*}[t]
\includegraphics[width=1.0\linewidth]{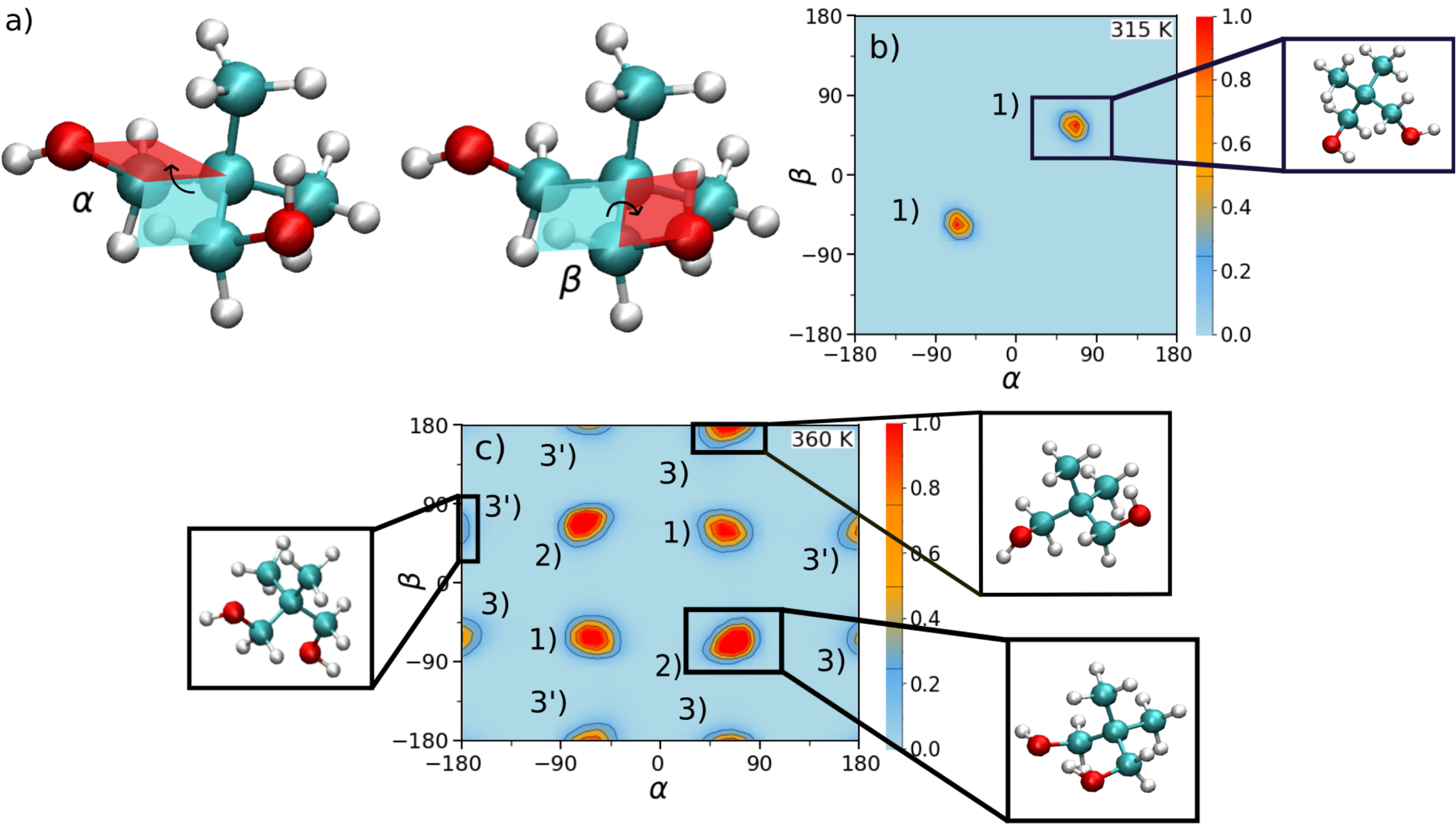}
       \caption{{\bf Molecular conformational degrees of freedom in the ordered and disordered phases of NPG.}
	(a)~Representation of the two dihedral angles describing different NPG molecular conformations. 
	Average dihedral angle histograms obtained from $NPT$--MD simulations for (b)~the low-$T$ ordered, and 
	(c)~high-$T$ disordered phases of NPG at zero pressure. Numbers in the figure indicate which high-probability
	spots describe equivalent or different molecular conformations.
       }
\label{fig5}
\end{figure*}

\subsection{Orientational degrees of freedom}
\label{subsec:ori}
In plastic crystals, orientational disorder resulting from molecular rotations emerges at high temperatures and 
typically is accompanied by a substantial volume increase (Fig.~\ref{fig1}a) \cite{lloveras19,li19,cazorla19b,aznar20}. 
In Sec.~\ref{sec:discussion}, we provide a detailed quantitative analysis of $S_{\rm ori}$ estimated across the 
OD phase transition. Here, we concentrate on the molecular orientational differences between the low-$T$ ordered 
and high-$T$ disordered phases of NPG.  

To study the orientation of the NPG molecules, which in this work are assumed to rotate independently of each 
other \cite{escorihuela24}, we define a co-mobile reference system for each molecule as follows (Fig.~\ref{fig4}a). 
The origin is placed at the C atom nearest to the molecular center of mass, which is exclusively bonded to other 
C atoms. The $z$-axis is defined along the direction connecting this central C atom to another C atom in a hydroxyl 
group (denoted as the molecular polar axis). The $x$-axis is chosen perpendicular to the plane containing the 
$z$-axis and the two C atoms closer to the reference system origin. Finally, the $y$-axis is determined as perpendicular 
to the $x$--$z$ plane (Fig.~\ref{fig4}a).  

As a molecule begins to rotate, its change in orientation relative to a fixed reference system (e.g., the co-mobile
reference system at the launch of the simulation) can be tracked by calculating the three angles $\lbrace \theta, 
\phi, \psi \rbrace$. These angles are illustrated in Fig.~\ref{fig4}a and their definition follow standard polar 
coordinates and Euler conventions \cite{pardo16,pardo15}. It is important to note that the choice of the fixed 
reference system is arbitrary, as it does not affect the resulting angular probability distribution diagrams (except 
for a uniform constant shift).

Figures~\ref{fig4}b--e present bidimensional histograms of the polar angle $\theta$ as a function of the rotational angles 
$\phi$ and $\psi$, which together completely define the orientation of NPG molecules. These histograms are shown for both 
the low-$T$ ordered (Figs.~\ref{fig4}b,d) and high-$T$ disordered (Figs.~\ref{fig4}c,e) phases. The averages were computed 
over $1,000$ system configurations extracted from the $NPT$--MD simulations at $0.5$~ps intervals, to minimize the correlations 
between the samples and improve the statistics. In the orientational maps, red regions indicate high-probability molecular 
orientations, while blue regions indicate low-probability molecular orientations.

As expected, the number of possible molecular orientations in the high-$T$ disordered phase is greater than in the 
low-$T$ ordered phase. In the ordered phase, four distinct reddish spots are observed in the $\phi$--$\cos{\theta}$ 
and $\psi$--$\cos{\theta}$ histograms. These spots correspond to the specific orientations of the four molecules within the 
monoclinic unit cell ($Z = 4$). In contrast, the disordered phase exhibits a more complex behavior. Specifically, the 
molecular polar axis adopts three distinct orientations, aligning approximately with the angles $\theta = 0^{\circ}$, 
$90^{\circ}$, and $180^{\circ}$. Furthermore, the preferred molecular orientations are threefold for apical configurations 
($\cos{\theta} = -1, 1$) and sixfold within the equatorial plane ($\cos{\theta} = 0$), totalling to $12$. Consistently, 
this number of preferred molecular orientations is equal to the number of nearest neighbours in the face-centered crystal 
arrengement of the high-$T$ disordered phase.  

An important observation is that, in the low-$T$ ordered phase, the high-probability orientation spots are isolated 
(Figs.~\ref{fig4}b,d). The absence of orientational paths connecting these spots effectively renders the probability 
of a molecule adopting a different orientation from its initial one to be zero. In contrast, the high-$T$ disordered 
phase displays a fundamentally different pattern. As shown in Figs.~\ref{fig4}c,e, the preferred molecular orientations 
are interconnected by non-zero probability paths, indicating that molecular orientations can evolve over time via 
transitions along these paths. These orientational transitions are not random but rather well-defined, unlike in an 
ideal molecular rotor, where all three-dimensional orientations are equally probable, and no specific preferred 
orientations or transition paths exist \cite{cazorla24}.

\begin{figure*}[t]
\includegraphics[width=1.0\linewidth]{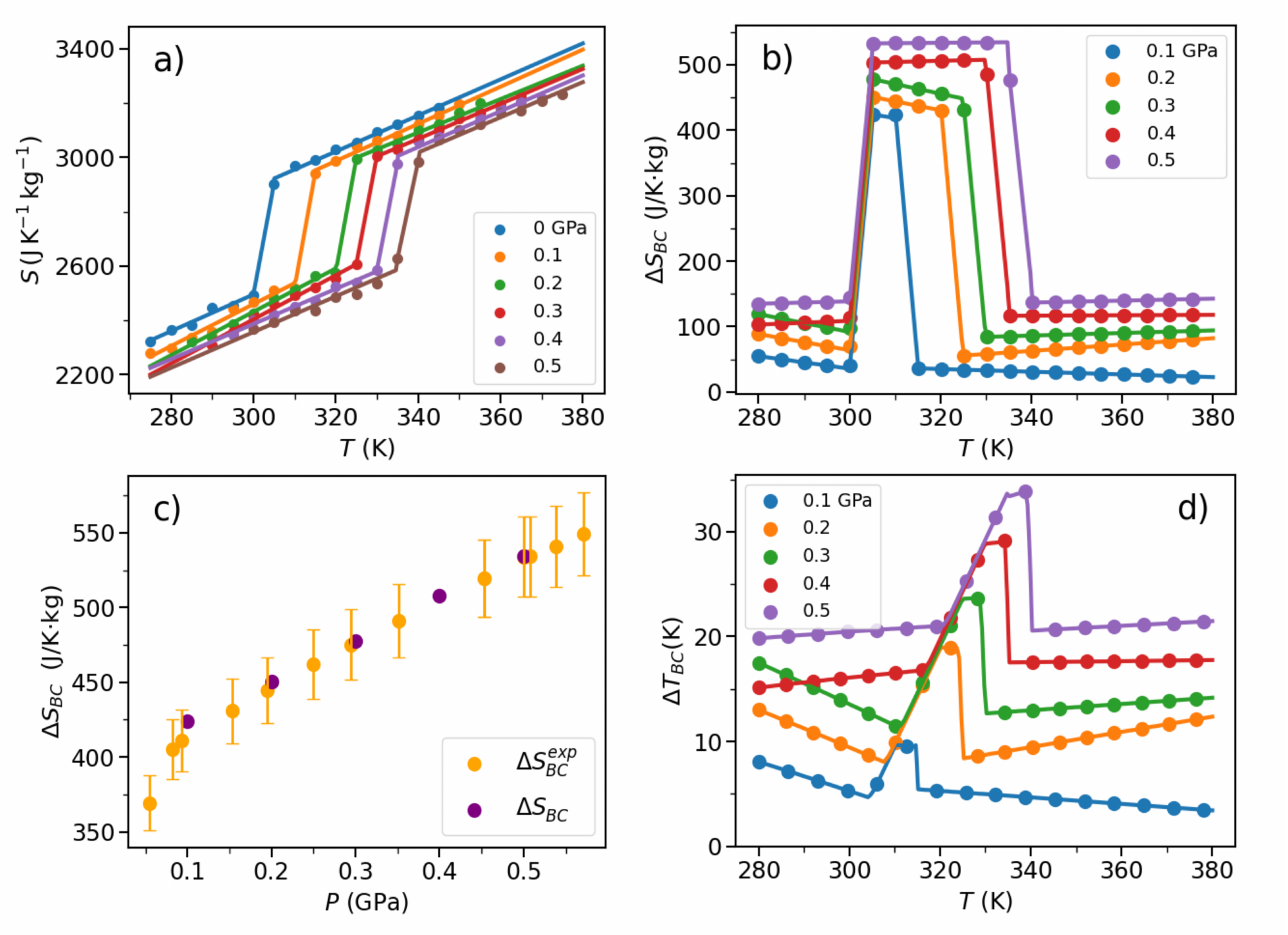}
       \caption{{\bf Estimated barocaloric performance of NPG.}
	(a)~Entropy curves expressed as a function of temperature at different fixed pressures.
	(b)~Estimated barocaloric isothermal entropy changes in NPG. 
	(c)~Comparison between experimental and theoretical barocaloric isothermal entropy changes.
	(d)~Estimated barocaloric adiabatic temperature changes in NPG.
	Solid lines are guides for the eye.
       }
\label{fig6}
\end{figure*}

\subsection{Conformational degrees of freedom}
\label{subsec:conf}
Molecules in plastic crystals exhibit a degree of flexibility, allowing them to undergo conformational changes 
driven by thermal excitations and intermolecular interactions, both in the low-$T$ ordered and high-$T$ disordered 
phases \cite{seo22,li21,seo24}. In Sec.~\ref{sec:discussion}, we provide a detailed quantitative analysis of the 
variation in conformational entropy ($S_{\rm conf}$) associated with the OD phase transition in NPG. In this section, 
we concentrate on the molecular conformational differences between the ordered and disordered phases.

A NPG molecule can undergo conformational changes through internal rotations around its C--C, C--O, and O--H bond 
axes. To identify and quantify the population of molecules in the most predominant conformations, we define two 
dihedral angles, $\alpha$ and $\beta$, which are determined by the intersection of two flat surfaces. The first 
dihedral angle, $\alpha$, is defined by two planes: one containing the oxygen atom of a hydroxyl group (O1), the 
carbon atom bonded to O1 (C1), and the carbon atom bonded to C1 (C2); and another containing the C1 and C2 atoms, 
along with the carbon atom bonded to the oxygen atom of the second hydroxyl group (C3) (Fig.~\ref{fig5}a). Similarly, 
the second dihedral angle, $\beta$, is defined by the plane containing the C1, C2, and C3 atoms, and the plane 
containing the C2 and C3 atoms along with the oxygen atom of the second hydroxyl group (i.e., the one closer to 
C3, Fig.~\ref{fig5}a).

A histogram of the two dihedral angles, $\alpha$ and $\beta$, provides a detailed characterization of molecular 
conformations in NPG, akin to the Ramachandran plots used in biochemistry for proteins \cite{ramachandran}. We 
generated conformational maps for both the low-$T$ ordered (Fig.~\ref{fig5}b) and high-$T$ disordered (Fig.~\ref{fig5}c) 
phases. In the ordered phase, two high-probability regions are evident, which could initially be interpreted as 
distinct molecular conformations. However, symmetry considerations reveal that these regions represent the same 
molecular conformation (Fig.~\ref{fig5}b), differing only by a rotational transformation.

In the high-$T$ disordered phase, the number of high-probability regions increases substantially, with eight distinct 
spots observed, indicating a broader variety of molecular conformations. Nonetheless, symmetry arguments again 
reduce the number of unique conformations, as some spots correspond to equivalent configurations differentiated 
only by rotation. In particular, the eight high-probability regions reduce to four distinct molecular conformations 
in the orientationally disordered phase (labels 1, 2, 3, and 3' in Figs.~\ref{fig5}b,c). 

It is worth noting that transitional paths connecting different conformations are absent for the high-$T$ disordered 
phase (Fig.~\ref{fig5}c), in contrast to the orientational diagrams presented in Figs.~\ref{fig4}c,e. This discrepancy 
arises because molecular conformational changes occur very rapidly, within a timescale shorter than $0.5$~ps (i.e., the 
time interval used in this study for configuration sampling), and consequently are not captured.

\subsection{Estimation of barocaloric effects in NPG}
\label{subsec:baro}
Using the computational approach detailed in Sec.~\ref{sec:computation} and the data presented in Figs.~\ref{fig3}--\ref{fig5}, 
we can estimate the total entropy of NPG as a function of temperature and pressure, $S(P,T)$, and particularly across its 
OD phase transition. Similar to quasi-direct calorimetry experiments \cite{aznar17,lloveras19,aznar20,li24}, the barocaloric 
(BC) isothermal entropy change, $\Delta S_{\rm BC}$, and adiabatic temperature change, $\Delta T_{\rm BC}$, can be directly 
inferred from these entropy curves \cite{rurali24,cazorla24}. Specifically, $\Delta S_{\rm BC}$ is calculated as $S(P,T) - 
S(0,T)$, while $\Delta T_{\rm BC}$ is given by $T_{0}(S,P) - T(S,0)$, where $T_{0}$ satisfies the condition $S(T_{0},P) = 
S(T,0)$. It is important to note that hysteresis effects --often a significant limitation for practical applications 
\cite{lloveras21,aznar20}-- are not considered in this study, as all the $NPT$--MD simulations were conducted under 
thermodynamic equilibrium conditions.  

Figure~\ref{fig6}a presents the $S(P,T)$ curves calculated for NPG as a function of pressure and temperature. 
The computed BC isothermal entropy changes (Fig.~\ref{fig6}b) accurately capture the expected increase in 
$\Delta S_{\rm BC}$ under increasingly larger pressure shifts. Notably, the quantitative agreement between our 
calculated $\Delta S_{\rm BC}$ values and the experimental data \cite{lloveras19} is excellent within the numerical 
uncertainties (Fig.~\ref{fig6}c). For instance, for a pressure shift of $0.1$~GPa, the calculated isothermal 
entropy change is $425$~J~K$^{-1}$~kg$^{-1}$, which aligns perfectly with the colossal experimental value of 
$412 \pm 25$~J~K$^{-1}$~kg$^{-1}$. This remarkable agreement extends across the entire range of investigated 
pressures. For example, at a pressure shift of $0.5$~GPa, both the calculated and experimental $\Delta S_{\rm BC}$ 
values converge at approximately $530$~J~K$^{-1}$~kg$^{-1}$ (Fig.~\ref{fig6}c).  

Figure~\ref{fig6}d illustrates the adiabatic temperature changes estimated for NPG, obtained directly from the 
$S(P,T)$ curves shown in Fig.~\ref{fig6}a. Consistent with the $\Delta S_{\rm BC}$ results, our $NPT$--MD simulations 
and thermodynamic data analysis accurately reproduce the increasing trend of $\Delta T_{\rm BC}$ under increasingly 
larger pressure shifts. For instance, the maximum $\Delta T_{\rm BC}$ evaluated for a pressure shift of $0.1$~GPa is 
$10$~K, while for $0.5$~GPa the maximum value reaches approximately $35$~K (Fig.~\ref{fig6}d). These estimations 
align closely with the corresponding experimental values of $11$~K and $40$~K \cite{lloveras19}, respectively.

\begin{figure*}[t]
\includegraphics[width=1.0\linewidth]{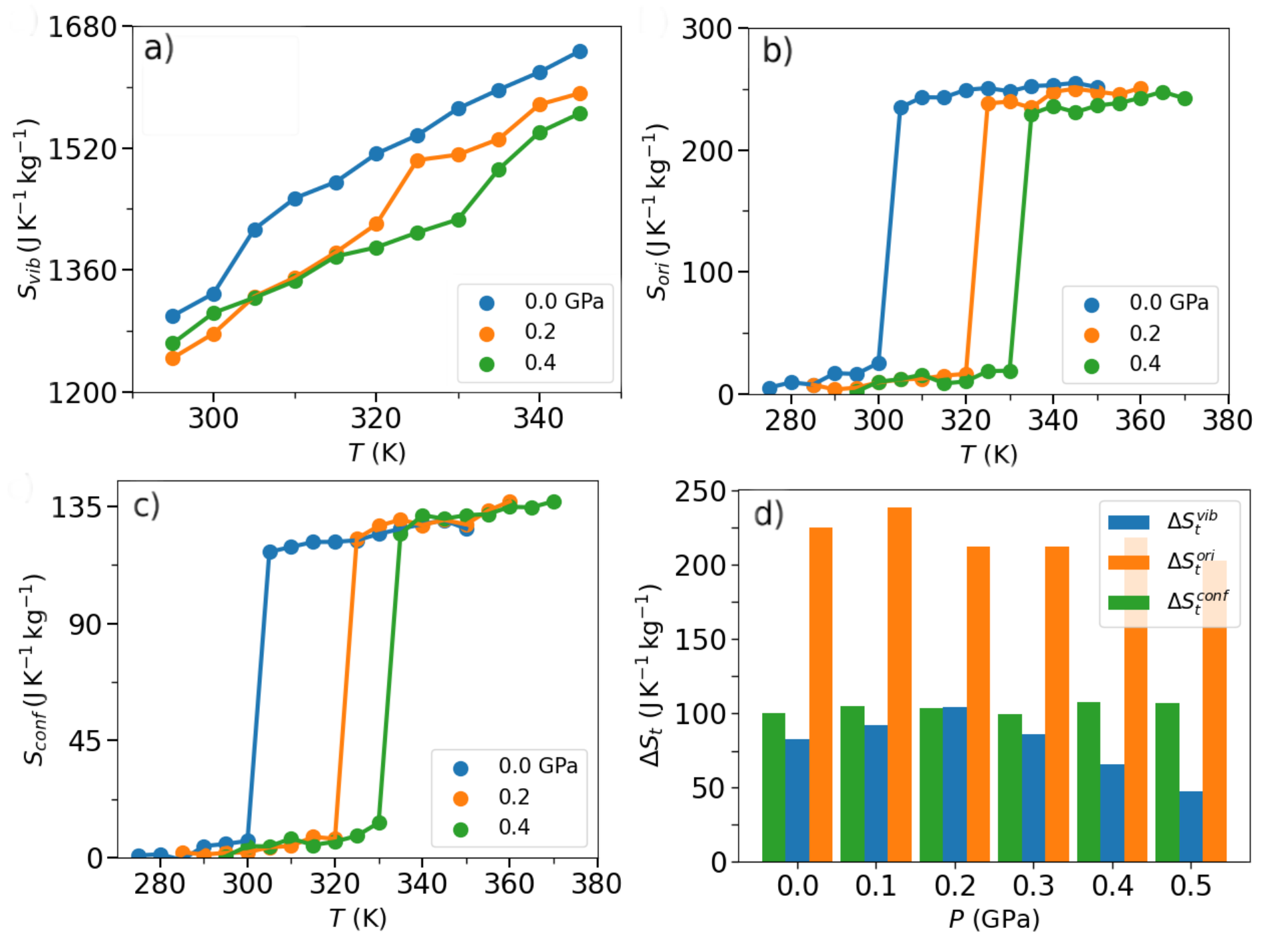}
       \caption{{\bf Vibrational, orientational, and conformational contributions to the entropy and transition 
	entropy change in NPG.} (a)~Vibrational, (b)~orientational and (c)~conformational entropies expressed as 
	a function of temperature and pressure. (d)~Contributions to the transition entropy change expressed as a 
	function of pressure. Solid lines are guides for the eye. 
       }
\label{fig7}
\end{figure*}

\section{Discussion}
\label{sec:discussion}
In the previous section, we have shown that the computational approach introduced in Sec.~\ref{sec:computation} 
combined with the results of our comprehensive $NPT$--MD simulations provide a quantitatively precise description 
of the OD phase transformation in NPG, whether driven by pressure or temperature, along with its distinctive BC 
performance as compared to experiments. To the best of our knowledge, this marks the first instance where a 
classical simulation study based on a force field --entirely free of \emph{ad hoc} free-energy model assumptions and 
empirical input-- has achieved such a high degree of accuracy and quantification. This unique capability allows us 
to evaluate and compare the individual contributions to the phase transition entropy change ($\Delta S_{t}$) and 
BC isothermal entropy change ($\Delta S_{\rm BC}$) in NPG, as we proceed to do next.

Figures~\ref{fig7}a--c illustrate the partial contributions to the total entropy --vibrational ($S_{\rm vib}$), orientational
($S_{\rm ori}$), and conformational ($S_{\rm conf}$)-- as functions of pressure and temperature. Together, these three 
contributions combine to produce the $S(P,T)$ curves shown in Fig.~\ref{fig6}a. In all cases, a significant increase in 
entropy is observed at the transition temperature corresponding to each pressure, $T_{t}(P)$. The vibrational entropy, 
$S_{\rm vib}$, increases monotonically with temperature and exhibits large values even far from the phase transition region. 
In contrast, the two angular entropies, $S_{\rm ori}$ and $S_{\rm conf}$, remain mostly negligible at temperatures below 
$T_{t}(P)$ and increase abruptly at the phase transition. Notably, the largest phase transition entropy change originates 
from $S_{\rm ori}$, which alone is already colossal (i.e., $\Delta S_{\rm ori} \geq 100$~J~K$^{-1}$~kg$^{-1}$). However, the 
combined contributions of $\Delta S_{\rm vib}$ and $\Delta S_{\rm conf}$ are comparable in magnitude to $\Delta S_{\rm ori}$, 
thus emphasizing their importance in the total entropy change.

For a pressure shift of $0.2$~GPa, the isothermal entropy change attributable solely to molecular reorientations, 
$\Delta S_{\rm BC}^{\rm ori}$, is approximately $240$~J~K$^{-1}$~kg$^{-1}$ (Fig.~\ref{fig7}b). Similarly, the isothermal 
entropy change arising from molecular conformational changes, $\Delta S_{\rm BC}^{\rm conf}$, is $110$~J~K$^{-1}$~kg$^{-1}$ 
(Fig.~\ref{fig7}c). Finally, the vibrational contribution to the total isothermal entropy change is $\Delta S_{\rm BC}^{\rm vib} 
= 100$~J~K$^{-1}$~kg$^{-1}$ (Fig.~\ref{fig7}a). Consequently, in relative terms, molecular reorientations account for 
approximately $53$\% of the BC response of NPG, while molecular conformations and vibrational contributions represent $25$\% 
and $22$\%, respectively. Analogous quantitative results are obtained for other pressure shifts (i.e., $0.4$~GPa in 
Figs.~\ref{fig7}a--c). 

Figure~\ref{fig7}d depicts the relative contributions of the different entropy components to $\Delta S_{t}$ as a function 
of compression. At zero pressure, the dominant contributor is $S_{\rm ori}$, accounting for approximately $56$\% of the 
total entropy change. Notably, the $S_{\rm conf}$ contribution, often overlooked in previous studies analyzing phase-transition 
entropy changes in plastic crystals \cite{hui22,li20,oliveira23,marin24}, constitutes a significant $25$\%. The vibrational 
degrees of freedom, $S_{\rm vib}$, contribute the remaining $19$\%. At the highest investigated pressure, $0.5$~GPa, the 
contributions remain largely consistent, with $S_{\rm ori}$ contributing $56$\%, $S_{\rm conf}$ increasing slightly to $29$\%, 
and $S_{\rm vib}$ decreasing to $15$\%. The value of these partial $\Delta S_{t}$ contributions are consistent with those 
reported above for $\Delta S_{\rm BC}$.   

It is noteworthy that a previous study provided a tentative estimation of the vibrational and angular (i.e., orientational 
and conformational together) contributions to $\Delta S_{t}$ for NPG, based on straightforward thermodynamic and molecular 
symmetry arguments \cite{lloveras19}. Using the experimental values for the OD phase-transition volume change, the isobaric 
thermal expansion coefficient, and the isothermal compressibility averaged over the ordered and disordered phases, the authors 
estimated $\Delta S_{t}^{\rm vib} \approx 60$~J~K$^{-1}$~kg$^{-1}$ at zero pressure. This value is approximately $25$\% smaller 
than the vibrational contribution reported in this work (i.e., $80$~J~K$^{-1}$~kg$^{-1}$, Fig.~\ref{fig7}d). Similarly, by 
considering up to $60$ distinct and equally probable configurations per NPG molecule in the disordered phase (i.e., $10$ 
molecular orientations $\times$ $6$ possible conformations) and one in the ordered phase, they estimated $\Delta S_{t}^{\rm ori 
+ conf} = M^{-1} R \ln(60) \approx 330$~J~K$^{-1}$~kg$^{-1}$, where $M = 104.15$~g~mol$^{-1}$ and $R$ is the universal gas 
constant. This value aligns closely with our numerical result of $320$~J~K$^{-1}$~kg$^{-1}$ (Fig.~\ref{fig7}d).  

%Nevertheless, it is important to note that the number of possible molecular configurations considered for the disordered 
%phase in \cite{lloveras19} differs from that determined in the present atomistic simulation study (e.g., $4$ molecular 
%conformations, Fig.~\ref{fig5}). Consequently, the excellent agreement between the two estimations is likely to be 
%coincidental. In fact, a similar angular entropy evaluation based on static molecular symmetry arguments has been shown 
%to fail in characterizing CH$_{3}$NH$_{3}$PbI$_{3}$ \cite{escorihuela24}. The limitations of such static approaches likely 
%stem from oversimplifications, such as the assumption of delta-like probabilities for all possible molecular configurations
%determined by symmetry and the neglection of orientational transition paths. Thus, while a rough estimation of the angular 
%entropy change may serve as a reasonable initial guess, it is unlikely to provide quantitatively accurate results. 

The primary conclusion of this classical simulation study, based on the results presented in Fig.~\ref{fig7}, is that across 
the OD phase transition in NPG, the combined entropy change from lattice vibrations and molecular conformations is comparable 
in magnitude to that resulting from molecular reorientations (approximately $45$\% and $55$\%, respectively). Notably, within 
the former contributions, molecular conformations exhibit the largest entropy change. These findings, obtained from the archetypal 
plastic crystal NPG, are likely to be broadly applicable to other similar compounds.  

Importantly, while hydrogen bonding --which directly affects molecular rotational dynamics in plastic crystals \cite{tamarit97}-- 
has traditionally been central to explaining the microscopic origins of the colossal BC effects observed in NPG and similar 
materials \cite{hui22,li20}, the findings of this study suggest an expanded perspective. Our atomistic simulations demonstrate 
that molecular structural and lattice vibrational contributions together play an equally significant role. These factors should 
therefore be integrated into the rational engineering of plastic materials for solid-state cooling. Furthermore, incorporating 
molecular conformational and vibrational degrees of freedom into phenomenological free-energy models \cite{oliveira23,marin24} 
appears to be also essential for achieving reliable descriptions of BC performances in plastic crystals.

\section{Conclusions}
\label{sec:conclusion}
In this study, we elucidate the molecular mechanisms underpinning the colossal barocaloric effects observed in the 
plastic crystal neopentyl glycol using molecular dynamics simulations and advanced entropy evaluation methods. Our 
results reveal that the entropy changes associated with molecular reorientations, lattice vibrations, and conformational 
degrees of freedom are all significant and collectively contribute to the barocaloric performance of NPG. Notably, we 
establish that the joint contributions from lattice vibrations and molecular conformations are comparable to those 
from molecular reorientations, challenging the conventional view that rotational dynamics alone dominate such effects.

Furthermore, our simulations demonstrate excellent agreement with experimental data, both in terms of phase-transition 
entropy changes and barocaloric performance metrics, validating the accuracy of our computational approach. These findings 
emphasize the need to incorporate vibrational and conformational contributions into the rational design and modeling 
of advanced barocaloric materials. By broadening the understanding of entropy mechanisms in plastic crystals, this work 
provides valuable insights for developing next-generation solid-state cooling technologies with improved efficiency and 
functionality.

\section*{Methods}
\label{sec:methods}

{\bf Molecular dynamics simulations.}~We used the GROMACS simulation code \cite{gromacs} to perform systematic classical 
molecular dynamics (MD) simulations in the $NPT$ ensemble for bulk NPG using a CHARMM-type force field \cite{charmm,swissparam1,
swissparam2}. The average temperature and pressure values were set using Nos\'{e}-Hoover thermostats and barostats with 
a mean fluctuation of $5$~K and $0.01$~GPa, respectively. The simulation box contained a total of $4,104$ atoms (equivalent 
to $216$ NPG molecules) and periodic boundary conditions were applied along the three Cartesian directions. The long-range 
electrostatic interactions were calculated by using a particle-particle particle-mesh solver to compute Ewald sums up to 
an accuracy of $10^{-4}$~kcal~mol$^{-1}$~\AA$^{-1}$ in the atomic forces. The cutoff distance for evaluation of the potential 
energy was $12$~\AA. 

To determine the phase-transition temperature of NPG under broad pressure conditions, we conducted comprehensive $NPT$--MD 
simulations in the temperature range $200 \le T \le 400$~K, taken at intervals of $5$~K. In our $NPT$--MD simulations, 
the temperature and pressure were steadily increased up to targeted values over $1$~ns. Subsequently, the system was 
equilibrated at the selected conditions for $1$~ns. The production runs then lasted for about $500$~ps using a time step 
of $2$~fs, from which the velocities of the atoms and other key quantities (e.g., the potential energy and volume of the 
system) were extracted. From the production $NPT$--MD runs, a total of $1,000$ equispaced configurations were retrieved to 
obtain uncorrelated structural data and generate accurate angular probability densities and histograms. 
\\

{\bf Molecular angular and entropy analysis.}~The NPG angular degrees of freedom have been retrieved from the atomic 
configurations generated during the $NPT$--MD simulations with the help of the in-house developed, freely available 
and open-source software ANGULA \cite{angula}. ANGULA is designed to automatically and unsupervisedly determine the 
angles defining the orientational structure of molecular disordered crystals from data files containing their atomic 
positions. Among its many capabilities, ANGULA can generate angular probability density maps and directional radial 
distribution functions directly from sequences of molecular configurations. In this study, the angular molecular entropy 
terms $S_{\rm ori}$ and $S_{\rm conf}$ have been directly computed from the outputs of our $NPT$--MD simulations with 
ANGULA \cite{angula}. 
\\

\section*{Acknowledgments}
The authors acknowledge financial support from MCIN/AEI/10.13039/501100011033 and ERDF/EU under the project 
PID2023-146623NB-I00 and through the ``María de Maeztu'' Program for Units of Excellence (CEX2023-001300-M), 
as well as from the Generalitat de Catalunya (Grant No. 2021SGR-00343). C.C. also acknowledges financial support 
from MCIN/AEI/10.13039/501100011033 under the fellowship RYC2018-024947-I and grant TED2021-130265B-C22. The 
authors also thankfully acknowledge technical support and computational resources at MareNostrum5 provided by 
Barcelona Supercomputing Center (FI-2023-1-0002, FI-2023-2-0004, FI-2023-3-0004, FI-2024-1-0005 and FI-2024-2-0003).
\\

\end{document}